\documentclass[prl,aps,showkeys,preprintnumbers,twocolumn,showpacs,amsmath,amssymb]{revtex4}
\usepackage{graphicx}
\usepackage{dcolumn}
\usepackage{bm}
\begin{document}
\preprint{\textit{Version October 11th 07}}
\title{Observation of extremely slow hole spin relaxation in self-assembled quantum dots}
\author{D. Heiss}
\author{S. Schaeck}
\author{H. Huebl}
\author{M. Bichler}
\author{G. Abstreiter}
\author{J. J. Finley}
\email{finley@wsi.tum.de}
\affiliation{Walter Schottky Institut, Technische Universit\"at M\"unchen, Am Coulombwall 3, D-85748 Garching, Germany}
\author{D. V. Bulaev}
\author{Daniel Loss}
\affiliation{Department of Physics and Astronomy, University of Basel, Klingelbergstrasse 82, CH-4056 Basel, Switzerland.}
\date{\today}
\begin{abstract}
We report the measurement of extremely slow \textit{hole} spin relaxation dynamics in small ensembles of self-assembled InGaAs quantum dots. Individual spin orientated holes are optically created in the lowest orbital state of each dot and read out after a defined storage time using spin memory devices. The resulting luminescence signal exhibits a pronounced polarization memory effect that vanishes for long storage times.  The hole spin relaxation dynamics are measured as a function of external magnetic field and lattice temperature. We show that hole spin relaxation can occur over remarkably long timescales in strongly confined quantum dots (up to $\sim$270~$\rm{\mu}$s), as predicted by recent theory.  Our findings are supported by calculations that reproduce both the observed magnetic field and temperature dependencies. The results suggest that hole spin relaxation in strongly confined quantum dots is due to spin orbit mediated phonon scattering between Zeeman levels, in marked contrast to higher dimensional nanostructures where it is limited by valence band mixing.  
\end{abstract}
\maketitle
The ability to manipulate and readout the spin of isolated charges in semiconductor quantum dots (QDs) has attracted considerable interest over recent years due to the strong potential for nanoscale spintronic devices and spin based quantum information processing.\cite{Awschalom02,LossDiVincenzo98} In this context, QDs are essential since their fully quantized electronic structure strongly inhibits spin relaxation mediated by scattering processes that couple to the spin via spin-orbit (SO) interactions.\cite{Khaetskii01,Golovach04,Coish06} Until now \textit{electron} spin dynamics have attracted by far the most attention, due to the rapid progress in the development of spin readout and control techniques \cite{Petta05,Koppens06,Amasha06} and since the theoretical understanding is more mature.\cite{Coish06} For both electrostatic and optically active QDs the electron spin relaxation has been found to be rather slow (100~$\rm{\mu}$s$\leq~T_1^e\leq$100~ms) when subject to a moderate magnetic field, and shown to proceed by SO-mediated single phonon scattering.\cite{Amasha06,Kroutvar04,Coish06} However, despite the long $T_1^e$ times it has now become clear that hyperfine coupling of the electron spin with the randomly fluctuating nuclear spin system results in rather rapid spin dephasing ($T_2^e\sim$10~ns), that can only be overcome using multi pulse coherent control methods or nuclear state narrowing. \cite{Stephanenko06, Klauser06, Greilich06, Petta05,Koppens06}\\
Unlike electrons, holes couple more weakly to the nuclear spins via the hyperfine contact interaction since they have $p$-like central cell symmetry \cite{Klauser06}. This may provide an attractive route towards \textit{hole spin} based applications free from the complications caused by the fluctuating nuclear spin system. However, the hole spin lifetime ($T_1^h$) in III-V semiconductor nanostructures is generally much shorter than $T_1^e$ due to SO-mixing of heavy (HH) and light hole (LH) valence bands.\cite{Lue05,Hilton02} This mixing is inhibited by motional quantization effects and enhanced hole spin lifetimes have been reported for quantum wells ($\sim100$ps - \cite{Damen91}), extending beyond $1$~ns when optically driven spin heating effects are avoided.\cite{Baylac95} For QD nanostructures, $T_1^h$ is expected to become even longer due to the combined effects of bi-axial compressive strain and motional quantization. These expectations have recently been supported by studies of negatively charged trions in InAs\cite{Laurent05} and CdSe\cite{Flissikowski03} QDs, which indicate that $T_1^h\geq10$~ns, limited by the timescale for radiative recombination of the trions.  Very recent calculations have indicated  that $T_1^h$ can become much longer for isolated holes\cite{Bulaev05}, even exceeding $T_1^e$ in the limit when the energy separation between HH and LH bands far exceeds the orbital quantization energy in the valence band.\cite{Lue05} Systematic measurements of the bare hole spin relaxation dynamics are urgently required as input for theory.\\
Here we report direct measurements of $T_1^h$ for single holes in small ensembles of self-assembled InGaAs QDs. Time and polarization resolved magneto-optical spectroscopy is employed to obtain $T_1^h$ as a function of static magnetic field and lattice temperature. Our results show that single hole spin relaxation can proceed over timescales \textit{comparable} to electrons in nominally identical QDs. Furthermore, we demonstrate that inter valence band SO-mixing does \textit{not} necessarily result in  for hole spin relaxation as is the case in higher dimensional systems. Calculations for $T_1^h$ due to SO-mediated single phonon scattering for QDs in which HH~-~LH mixing is negligible produce very good quantitative agreement\cite {Bulaev05} with our results, indicating that hole spin relaxation in self-assembled QDs is mediated by the same mechanism as for electrons.\\
\begin{figure}[t]
    \begin{center}
        \includegraphics[width=\columnwidth]{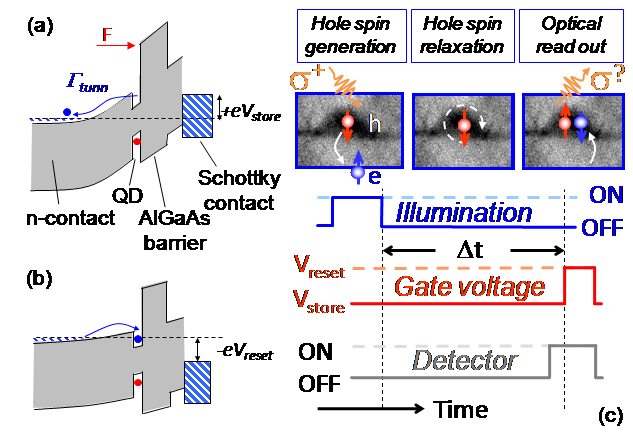}
    \end{center}
    \caption{(color online) Schematic band profile of the devices investigated in the spin generation (a) and readout (b) phases of the time gated EL measurement. (c) The temporal sequence of optical and electrical control pulses during the measurement and detector gating.}
\end{figure}
%
The devices investigated are $n-i-$metal (Al)GaAs Schottky photodiodes containing a single layer of self-assembled In$_{0.5}$Ga$_{0.5}$As QDs embedded within the $d=$110~nm thick intrinsic region.\cite{Kroutvar03,Ducommun04,Kroutvar04} The schematic band profile and operating principles of the structures are summarized in Fig. 1(a)-(b). In the spin \emph{generation} mode (Fig. 1(a)) a reverse bias ($V= -V_{store}$) is applied to the gate with respect to the $n-$contact. Single, charge neutral excitons ($X^0$) are generated resonantly in the lowest orbital state of the dots. Due to the large axial electric field ($F_{store}\approx(-V_{store}-V_{bi})/d\sim$70~kV/cm) electrons readily tunnel out of the dots whilst holes remain stored by virtue of a 50~nm thick Al$_{0.45}$Ga$_{0.55}$As barrier immediately above the QD layer (Fig. 1). Further excitation of the same QD is prevented by a renormalization of the absorption energy by a few meV after the first charging event, ensuring the storage of only a single hole per QD.\cite{Kroutvar03}\\
The spin orientation of the optically generated holes is defined by the circular polarization of the excitation laser.\cite{Kroutvar03,Kroutvar04} Due to the combined effects of motional quantization and bi-axial compressive strain $X^0$ has predominant heavy hole character and is formed from single particle basis states with spin projections $S_{e,z}=\pm\frac{1}{2}$ and $J_{h,z}=\pm\frac{3}{2}$ along the strong quantization axis of the dots ($z$). These states combine to produce a pair of bright excitons with $J_z^{ex}=S_{e,z}+J_{h,z}$=$+1$ and $-1$, that are split in a magnetic field ($\vec{B}||\widehat{z}$) by the Zeeman energy ($E_{Z}^{ex}=(g_e^z-g_h^z)\mu_B B_z$). Mixing of these pure spin states due to the anisotropic $e-h$ exchange interaction is slow compared with the electron tunneling rate ($\Gamma_{tunn}\geq$100~GHz vs. $(\delta_1/\hbar)\leq$10~GHz) and, thus, excitation with $\sigma^{+}$ or $\sigma^{-}$ polarized light selectively creates spin \emph{up} ($+\frac{3}{2}$) or \emph{down} ($-\frac{3}{2}$) holes, respectively.\\
After optical charge generation and spin orientation the excitation laser is switched off and the hole spins are stored for a time $\Delta t$, during which they can interact with their environment. The distribution and spin orientation of the holes are then tested by applying a forward bias pulse ($V = +V_{reset}$). A diffusion current of electrons then flows into the QDs whereupon they recombine with the stored holes (Fig. 1(b)).\cite{comment} During reset, the current is chosen to be sufficiently high to compensate all the stored holes and therefore, no accumulative effects over several cycles of the experiment can take place. Furthermore, since the injected electrons are not spin polarized, the degree of circular polarization of the resulting, time delayed electroluminescence (EL) provides a measure of the spin projection of the ensemble of stored holes a time $\Delta t$ after generation.\cite{Pryor03}\\
Measurements were performed using 200~ns duration laser pulses delivered by a diode laser ($\hbar\omega_{laser}\sim$1.252~eV, $P_{exc}\sim $100~W/cm$^{2}$). These circularly polarized write pulses were synchronized with electrical control and readout pulses at a repetition frequency $f_{rep}\sim(\Delta t + $1${\mu}$s$)^{-1}$ as depicted schematically in Fig. 1(c). In order to ensure efficient suppression of scattered laser light during readout, a single photon detector was gated \emph{on} for $\sim$70~ns immediately prior to the readout voltage pulse (Fig. 1(c)). The spin storage time ($\Delta t$) is defined as the time delay between switching off the optical write pulse and the rising edge of the electrical readout pulse.
\\
\begin{figure}[h]
    \begin{center}
        \includegraphics[width=\columnwidth]{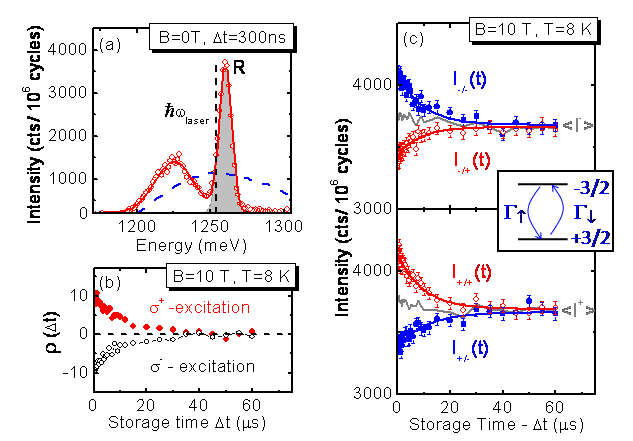}
    \end{center}
    \caption{ (color online) (a) Comparison of a non-resonantly excited PL spectrum (dashed line) with a hole \emph{storage} EL spectrum ($\hbar\omega_{laser}\sim$1.252~eV, $\Delta t=300$~ns at $f_{rep}\sim$200~kHz). (b) Temporal evolution of the peak intensity of $R$ recorded with circularly polarized excitation and detection ($I_{n/m}(\Delta t)$), where $n,m\in\{+,-\}$ denote the helicity of the excitation and detection, respectively. The dotted line shows the intensity $\langle~I\rangle$ averaged over the two polarizations. (c) Temporal evolution of the degree of circular polarization.
}
\end{figure}
A typical $B=$0~T hole storage EL spectrum is presented in Fig. 2(a) for a storage time of $\Delta t=$300~ns. The most prominent feature, labeled $R$ in Fig. 2(a), appears Stark shifted by $\sim6$~meV from $\hbar\omega_{ex}$ and arises from resonant optical storage of holes within the sub-ensemble of dots addressed by the excitation source.\cite{Kroutvar03} This identification is confirmed by the observation that peak $R$ tracks $\hbar\omega_{ex}$ as it is tuned through the inhomogeneously broadened QD absorption spectrum.\cite{Kroutvar03} The large Stark shift originates from a permanent excitonic dipole moment $|p_o| = 26\cdot 10^{-29}$~C/m, determined by measuring the spectral position of the storage peak as a function of $V_{store}-V_{reset}$.\cite{kroutvar_thesis} 
\\
Fig. 2(c) shows the temporal evolution of the peak intensity of $R$ at $B=10$~T recorded with circularly polarized excitation and detection for storage times in the range 0.3~$\rm{\mu}$s$\leq\Delta t\leq$60~$\rm{\mu}$s. Here, $I_{+/+}$ denotes the intensity of peak $R$ recorded with $\sigma^+$ polarization in both excitation and detection channels, the other three curves completing the four polarization permutations.  
The full dark gray lines on Fig. 2(b) show the storage EL intensities $\langle I^+\rangle$ and $\langle I^-\rangle$ for the corresponding excitation polarization \emph{averaged} over the two detection polarizations. Within experimental error $\langle I^+\rangle$ and $\langle I^-\rangle$ are independent of $\Delta t$ confirming that holes are not thermally redistributed between the dots via wetting layer states during the storage time.\cite{Ducommun04} We ensured that this condition applied to the range of $\Delta t$ investigated in this paper.\\
Careful examination of the data presented in Fig. 2(c) indicates the presence of a significant \emph{polarization memory} effect; the storage EL is co-polarized with the excitation laser for short storage times ($\Delta t\ll40\mu$s) with a degree of circular polarization $|\rho_h(t)|=(I_{\sigma^+}(t)-I_{\sigma^-}(t))/(I_{\sigma^+}(t)+I_{\sigma^-}(t))$ up to $10\%$ as shown in Fig. 2(b). As $\Delta t$ increases $|\rho_h|$ decreases exponentially for both $\sigma^+$ or $\sigma^-$ excitation, reaching an equilibrium value $\rho_h=0\pm2\%$ for long storage times ($\Delta t\geq40\mu$s).\\
To extract quantitative information from these data we use rate equations applicable to a non-interacting ensemble of $N_{tot}$ two-level systems to represent the total population of $+\frac{3}{2}$ and $-\frac{3}{2}$ holes in the QDs probed by our experiment (Fig. 2(b)(inset)). The rates for up-scattering ($-\frac{3}{2} \rightarrow +\frac{3}{2}$) and down-scattering ($+\frac{3}{2} \rightarrow -\frac{3}{2}$) are represented by $\Gamma_{\uparrow}$ and $\Gamma_{\downarrow}$, respectively. Using this model the population of a specific spin species, a time $t$ after generation, is given by  
\begin{equation}
N_{\uparrow(\downarrow)}(t)=(N_{\uparrow(\downarrow)}(0)-N_{\uparrow(\downarrow)}^{equ})\exp\left(-\frac{t}{T_1^h}\right)+N_{\uparrow(\downarrow)}^{equ}\\
\label{eqn1}
\end{equation}
where $N_{\uparrow(\downarrow)}^{equ}=\frac{\Gamma_{\downarrow(\uparrow)}}{\Gamma_{\uparrow}+\Gamma{\downarrow}}\cdot N_{tot}$ is the Boltzmann distribution at thermal equilibrium, $T_1^h=1/(\Gamma_{\uparrow}+\Gamma_{\downarrow})$ is the hole spin lifetime and $N_{\uparrow(\downarrow)}(0)$ the initial population of spin up (down) holes at $\Delta t=$0~s. Since the intensity of the storage luminescence a time $\Delta t$ after generation is proportional to $N_{\uparrow(\downarrow)}$ the curves presented in Fig. 2(b) ($I(t)$) can each be fit by
\begin{equation}
I_{n/m}(t)=\left(I_{n/m}(0)-\langle I^n\rangle\right)\exp\left(-\frac{t}{T_1^h}\right)+\langle I^n\rangle
\label{eqn2}
\end{equation}
where $I(0)$ is the corresponding intensity at $\Delta t=$0 and $\langle I\rangle=\langle I^+\rangle=\langle I^-\rangle$ is the saturation value for long storage times. The best fits to all four decay curves are presented as full lines on Fig. 2(b). Most importantly, using this model all four traces are well described using a \emph{single} value for the hole spin lifetime of $T_1^h\sim$11$\pm$3~$\rm{\mu}$s. No indications of a multi exponential decay were found for timescales longer than our temporal resolution limit ($\sim$200~ns), an observation which indicates that our measurement probes a single relaxation mechanism, free from inhomogeneous ensemble effects. Furthermore, any competing relaxation mechanisms which operate over the timescales studied can be excluded.\\
The observation of such remarkably slow hole spin relaxation dynamics in QDs contrasts strongly with bulk III-V materials where spin flip scattering occurs over sub picosecond timescales due to SO-mixing of the valence band spin states close to $k=0$.\cite{Hilton02} This mixing is partially inhibited by motional quantization effects in quantum wells, leading to enhanced hole spin lifetimes longer than $\sim$1~ns at low temperatures.\cite{Baylac95} In contrast, the fully quantized electronic structure in QDs restricts the phase space for such quasi-elastic scattering processes and results in much longer hole spin lifetimes in the present experiment.\cite{Bulaev05}\\
Inspection of Fig. 2(b) shows that even for storage times $\Delta t \ll T_1^h$, $|\rho_h|$ is limited to $\sim 10\%$, much lower than for \textit{electron} spin memory devices where $|\rho_e| \geq$60\% is commonly observed.\cite{Kroutvar04} 
At long storage times ($\Delta t \gg T_1^h$) the degree of circular polarization of the storage luminescence tends toward zero. Here, $|\rho_h|$ is similarly lower than the expected value of $|\rho_h|=$24\% according to Boltzmann statistics.\cite{gFactor} 
The origin of this low degree of polarization is not yet fully clear. However, we note that a reduction of $|\rho_h|$ may arise during the readout phase of the measurement, where the system is most strongly disturbed by switching the electric field and injection of an electron current to the dots. We note that any such perturbation of the spin orientation during readout has \emph{no influence} on dynamics probed in the storage phase of the present experiment, influencing only the magnitude of $|\rho_h|$. In contrast, a fast partial decay of the hole spin orientation over timescales faster than our temporal resolution ($\sim$200~ns) cannot be absolutely excluded as source for the reduced $|\rho_h|$. However, we note that no evidence for multi-exponential dynamics was found, indicating that our measurements probe a single relaxation mechanism, and previous work has already shown that $T_1^h\geq10$~ns\cite{Flissikowski03,Laurent05}, suggesting that we would see evidence for any fast relaxation in our experiment.\\
We continue by discussing the dependence of $T_1^h$ on external magnetic field $B$ along the growth direction and lattice temperature $T$. Fig. 3(a) shows the excess degree of circular polarization following $\sigma^+$ excitation and detection, $\left(I_{+/+}-\langle I\rangle\right)$, plotted as a function of $\Delta t$ for $B=$5-11~T at $T=$8~K. In all cases the time dependence of the polarization is well described by a mono exponential decay (Eq. \ref{eqn2}), with a time constant that becomes longer as the $B$-field is reduced. The values of $T_1^h$ obtained are presented in Fig. 3(b) and directly compared with equivalent data recorded from an electron spin memory device measured under the same experimental conditions.\cite{Kroutvar03,Kroutvar04,Heiss05} As the B-field is reduced from 12-1.5~T the hole spin lifetime is found to increase from 8$\pm$3 to 270$\pm$180~$\rm{\mu}$s. The spin relaxation time measured at $B=$1.5~T is more than four orders of magnitude longer than recent reports for single CdSe dots where a lower limit of $\sim$10~ns was measured.\cite{Flissikowski03} Furthermore, the ratio $T_1^e/T_1^h$ lies in the range $\sim$5-10 over the whole range of $B$-field investigated, in strong contrast with quantum wells where $T_1^e/T_1^h$ is typically $\geq$10$^3$ due to SO-mixing of HH and LH valence bands. When such SO-valence band mixing is suppressed by motional quantization we expect hole spin relaxation to be dominated by SO-mediated spin-lattice coupling, as is the case for electron spin relaxation in self-assembled QDs. To test this hypothesis we studied the temperature dependence of $T_1^h$. Fig. 3(c) shows the temperature dependence of $T_1^h$ for magnetic fields of $B=$6~T and 10~T. For both magnetic fields a very clear $T_1^{h}\propto T^{-1}$ dependence is observed, exactly as was found previously for electron spin relaxation in self-assembled QDs.\cite{linearTEMP,Heiss05,Kroutvar04,Kroutvar03}
The observation of mono-exponential decays and comparable spin lifetimes for electrons and holes combined with $T^{-1}$ temperature and strong $B$-field dependencies strongly indicates that the dominant relaxation mechanism for hole spins is, indeed, due to SO mediated spin-lattice coupling as has been shown to be the case for electrons.\cite{Kroutvar04,Amasha06}\\
To further support this conclusion we calculated the heavy hole spin relaxation time in disc like QDs having strong motional quantization along the \textit{z}-direction ($\hbar\omega_z$) and much weaker in-plane confinement ($\hbar\omega_{x,y}$).\cite{Bulaev06} For such a QD-geometry the energy splitting between discrete states derived from HH and LH valence bands is much larger than the orbital quantization energy and the valence band orbital states were, therefore, assumed to have pure HH character. Only spin relaxation mediated by single phonon scattering processes were included and the calculations were performed pertubatively using Bloch-Redfield theory to describe the spin motion of the system.(see Eq. (7) in Ref. \cite{Bulaev05}) The best fit to our experimental data is presented as the full lines on Fig. 3(b) and Fig. 3(c) for the B-field and temperature dependencies, respectively, obtained using the following parameters $\hbar\omega_{x,y}=$6~meV, $\hbar\omega_{z}=$100~meV, $g_h^z=$-0.6 and $m_h^*=0.03m_0$. This parameter set provides a good description of $T_1^h(B,T)$, providing strong support for our hypothesis; that hole spin relaxation in self-assembled QDs is mediated by the same mechanism as for electrons, namely spin-phonon scattering between the Zeeman levels.
%
\begin{figure}[h]
    \begin{center}
        \includegraphics[width=\columnwidth]{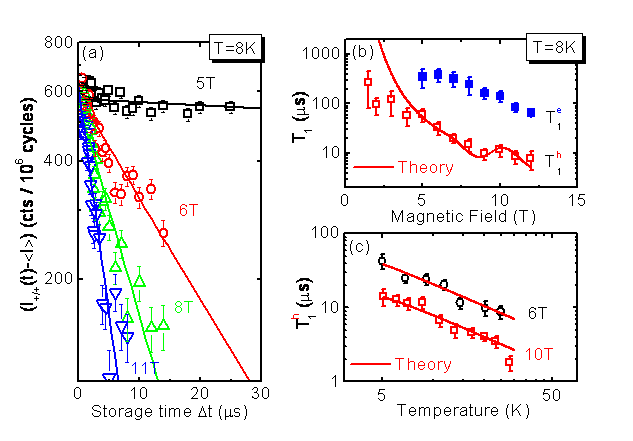}
    \end{center}
    \caption{ (a) Temporal dependence of the excess circular polarization for $B$ ranging from 5-11~T. (b) Comparison of the B-field dependence of $T_1^h$ (open symbols) and $T_1^e$ (filled symbols) for identical QD-material and experimental conditions. The filled line shows the calculated hole spin relaxation time using the model of ref. \cite{Bulaev06}. (c) Comparison of measured (open symbols) and calculated (full lines) dependence of $T_1^h$ on lattice temperature.}
\end{figure}
\\
In summary, we have measured extremely slow spin relaxation dynamics for holes in small ensembles of self assembled InGaAs QDs. Systematic investigations of $T_1^h$ as a function of magnetic field and temperature suggest that the hole spin relaxation proceeds by SO-mediated single phonon scattering, unlike the situation for hole spins in higher dimensional nanostructures. This conclusion is supported by good quantitative agreement with theoretical calculations of $T_1^h$ due to phonon mediated hole spin relaxation.
\\

The authors gratefully acknowledge financial support by the \textit{DFG} via \textit{SFB} 631 and the German excellence initiative via \textit{Nanosystems Initiative Munich} (NIM) and by the Swiss NF and the $NCCR$ Nanoscience Basel.


\begin{thebibliography}
\bibitem{comment}  We note that optical spin heating effects, similar to those reported in luminescence experiments on quantum wells\cite{Baylac95}, cannot occur in the presently reported experiment since optical hole generation occurs over timescales that are much shorter than the hole spin relaxation times ($<1$~ns c.f. $\sim8-270\mu$~s) and all stored charge is removed from the quantum dots after each cycle of the measurement, preventing accumulative effects, including spin heating.  Typical measurements average over several million equivalent hole spin $write$-$store$-$read$ cycles.
\bibitem {linearTEMP} The inverse temperature dependence arises since $T_1^{h -1}\propto(N(E_Z)+1)$, where $N_Z={( \exp(-g_h\mu_{B}B/k_{B}T}-1)^{-1}$ is the Bose-Einstein factor. In the high temperature regime when $k_{B}T\gg-g_h\mu_{B}B$,the Bose-Einstein factor can be approximated by $N_Z\approx{-g_h\mu_{B}B/k_{B}T}$ and we obtain $T_{1}^h \propto T^{-1}$.
\end{thebibliography}
\end{document}